\documentclass[aps,a4paper,superscriptaddress,showpacs,preprintnumbers,amsmath,amssymb]{revtex4}

\usepackage{ulem}

\usepackage{psfrag} \usepackage{graphicx} \usepackage{dcolumn}
\usepackage{color} \usepackage{latexsym,amsfonts} \usepackage{bm}
\usepackage{amssymb}
\begin{document}

\title{Infrared Properties of Hadronic Structure of Nucleon\\ in
  Neutron Beta Decays to Order $O(\alpha/\pi)$ in Standard $V - A$
  Effective Theory\\ with QED and Linear Sigma Model of Strong
  Low--Energy Interactions}

\author{A. N. Ivanov}\email{ivanov@kph.tuwien.ac.at}
\affiliation{Atominstitut, Technische Universit\"at Wien, Stadionallee
  2, A-1020 Wien, Austria}
\author{R.~H\"ollwieser}\email{roman.hoellwieser@gmail.com}
\affiliation{Atominstitut, Technische Universit\"at Wien, Stadionallee
  2, A-1020 Wien, Austria}\affiliation{Department of Physics,
  Bergische Universit\"at Wuppertal, Gaussstr. 20, D-42119 Wuppertal,
  Germany} \author{N. I. Troitskaya}\email{natroitskaya@yandex.ru}
\affiliation{Atominstitut, Technische Universit\"at Wien, Stadionallee
  2, A-1020 Wien, Austria}
\author{M. Wellenzohn}\email{max.wellenzohn@gmail.com}
\affiliation{Atominstitut, Technische Universit\"at Wien, Stadionallee
  2, A-1020 Wien, Austria} \affiliation{FH Campus Wien, University of
  Applied Sciences, Favoritenstra\ss e 226, 1100 Wien, Austria}
\author{Ya. A. Berdnikov}\email{berdnikov@spbstu.ru}\affiliation{Peter
  the Great St. Petersburg Polytechnic University, Polytechnicheskaya
  29, 195251, Russian Federation}

\date{\today}

\begin{abstract}
Within the standard $V - A$ theory of weak interactions, Quantum
Electrodynamics (QED) and the linear $\sigma$--model (L$\sigma$M) of
strong low--energy hadronic interactions we analyse infrared
properties of hadronic structure of the neutron and proton in the
neutron $\beta^-$--decays to leading order in the large nucleon mass
expansion. We confirm validity and high confidence level of
contributions of hadronic structure of the nucleon to the radiative
corrections, calculated by Sirlin (Phys. Rev. {\bf 164}, 1767 (1967))
to leading order in the large nucleon mass expansion. At the level of
order $10^{-5}$ relative to Sirlin's infrared divergent contribution
to the neutron radiative $\beta^-$--decay (inner bremsstrahlung) we
find an infrared divergent contribution, induced by hadronic structure
of the nucleon through the one--pion--pole exchange, to the rate of
the neutron lifetime from the neutron radiative $\beta^-$--decay,
which should be cancelled by contributions of virtual photon exchanges
to the neutron $\beta^-$--decay. Following Ivanov {\it et al.}
1805.09702 [hep-ph] we argue that a consistent analysis of such a
cancellation may be carried out well in the combined quantum field
theory including the Standard Electroweak Model (SEM) and the
L$\sigma$M of strong low--energy interactions, where the effective $V
- A$ hadron--lepton current--current vertex is caused by the
$W^-$--electroweak--boson exchange.
\end{abstract}
\maketitle

\begin{figure}
\centering \includegraphics[height=0.12\textheight]{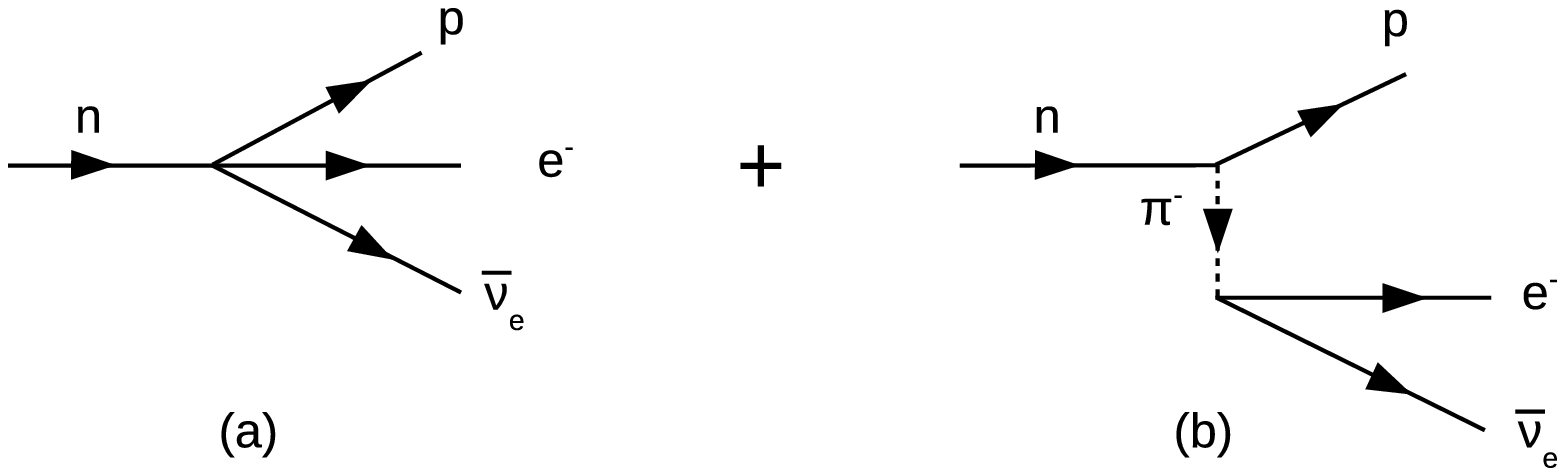}
  \caption{The Feynman diagrams, describing the amplitude of the
    neutron $\beta^-$--decay, defined in the limit $m_{\sigma} \to
    \infty$, to leading order in the large nucleon mass expansion and
    after renormalization in the L$\sigma$M.}
\label{fig:fig1}
\end{figure}

It is well--known that the neutron radiative $\beta^-$--decay $n \to p
+ e^- + \bar{\nu}_e + \gamma$ (inner bremsstrahlung) plays an
important for cancellation of infrared divergences, caused by virtual
photon exchanges, to the neutron $\beta^-$--decay to order
$O(\alpha/\pi)$ \cite{Berman1958}-\cite{Ivanov2017a}, where $\alpha$
is the fine--structure constant \cite{PDG2018}. As a physical process
the neutron radiative $\beta^-$--decay has been investigated
theoretically to order $O(\alpha/\pi)$ in
\cite{Gaponov1996}--\cite{Ivanov2017} (see also \cite{Ivanov2013}) and
to order $O(\alpha^2/\pi^2)$ in
\cite{Gardner2012,Gardner2013,Ivanov2017b}, respectively, and
experimentally in \cite{Nico2006}--\cite{Bales2016}. Recently
\cite{Ivanov2018b} we have analysed gauge properties of hadronic
structure of the nucleon in the neutron $\beta^-$--decays within the
combined quantum field theory including the standard $V - A$ effective
theory of weak interactions \cite{Feynman1958}--\cite{Marshak1969},
Quantum Electrodynamics (QED) and the linear $\sigma$--model (the
L$\sigma$M) of strong low--energy interactions
\cite{GellMann1960}--\cite{DeAlfaro1973}, which is renormalizable
\cite{Bernstein1960,Lee1969,Gervais1969,Mignaco1971,Strubbe1972} and
in the infinite limit of the scalar $\sigma$--meson mass $m_{\sigma}
\to \infty$ reproduces the results of the current algebra
\cite{Weinberg1967,Gasiorowicz1969}. We have shown that in the limit
$m_{\sigma} \to \infty$, to leading order in the large nucleon mass
expansion and after renormalization such a combined quantum field
theory defines the standard Lorentz structure of the matrix element of
the hadronic $n \to p$ transition of the neutron $\beta^-$--decay
including vector, axial--vector and pseudoscalar terms
\cite{Leitner2006,Ivanov2018}, where the contributions of strong
low--energy interactions are defined by the axial coupling constant
$g_A$ and the one--pion--pole exchange, and the gauge invariant
amplitude of the neutron radiative $\beta^-$--decay, where the
contributions of strong low--energy interactions are presented in
terms of the axial coupling constant $g_A$ and one--pion-- and
two--pion--pole exchanges. The Feynman diagrams of the amplitude of
the neutron $\beta^-$--decay, calculated to leading order in the large
nucleon mass expansion, are shown in Fig.\,\ref{fig:fig1}.
The amplitude of the neutron $\beta^-$--decay is defined by
\cite{Ivanov2018b}
\begin{eqnarray}\label{eq:1}
\hspace{-0.15in}M(n \to p e^-\bar{\nu}_e) = - G_V\langle
p(\vec{k}_p,\sigma_p)|J^+_{\mu}(0)|n(\vec{k}_n, \sigma_n)\rangle\,
\Big[\bar{u}_e\big(\vec{k}_e, \sigma_e\big) \gamma^{\mu}\big(1 -
  \gamma^5\big) v_{\nu}\big(\vec{k}_{\nu}, + \frac{1}{2}\big)\Big],
\end{eqnarray}
where $G_V = G_FV_{ud}/\sqrt{2}$ is the vector weak coupling constant,
and $G_F$ and $V_{ud}$ are the Fermi weak coupling constant and the
matrix element of the Cabibbo--Kobayashi--Maskawa (CKM) mixing matrix
\cite{PDG2018}, respectively, and $\bar{u}_e \gamma^{\mu}\big(1 -
\gamma^5\big) v_{\nu}$ is the matrix element of the leptonic $V - A$
current. The matrix element of  the hadronic  $n \to p$ transition
calculated in the limit $m_{\sigma} \to \infty$, to leading order in the
large nucleon mass expansion and after renormalization, takes the form
\cite{Ivanov2018b}
\begin{eqnarray}\label{eq:2}
\langle p(\vec{k}_p,\sigma_p)|J^+_{\mu}(0)|n(\vec{k}_n,
\sigma_n)\rangle = \bar{u}_p\big(\vec{k}_p,
\sigma_p\big)\Big\{\gamma_{\mu}\big(1 - g_A\gamma^5\big) +
\frac{\kappa}{2 m_N}\, i \sigma_{\mu\nu}q^{\nu} - \frac{2\,g_A m_N
}{m^2_{\pi} - q^2}\, q_{\mu} \gamma^5 \Big\}\,u_n\big(\vec{k}_n,
\sigma_n\big),
\end{eqnarray}
where the contributions of strong low--energy interactions are
presented by the axial coupling constant $g_A$ and the one--pion--pole
exchange.  The contribution of the one--pion--pole exchange is
necessary for local conservation of the axial hadronic current in the
limit $m_{\pi} \to 0$ \cite{Feynman1958,Nambu1960}
\begin{eqnarray}\label{eq:3}
\lim_{m_{\pi} \to\, 0}q^{\mu} \langle
p(\vec{k}_p,\sigma_p)|J^+_{\mu}(0)|n(\vec{k}_n, \sigma_n)\rangle = 0.
\end{eqnarray}
The one--pion--pole exchange contribution appears also in the current
algebra approach \cite{Marshak1969} (see also \cite{Adler1968}). The
term with the Lorentz structure $i\sigma_{\mu\nu}q^{\nu}/2m_N$, where
$m_N$ is the nucleon mass, describes the contribution of the weak
magnetism \cite{Bilenky1959,Wilkinson1982} with the isovector
anomalous magnetic moment of the nucleon $\kappa$.

\begin{figure}
\centering \includegraphics[height=0.23\textheight]{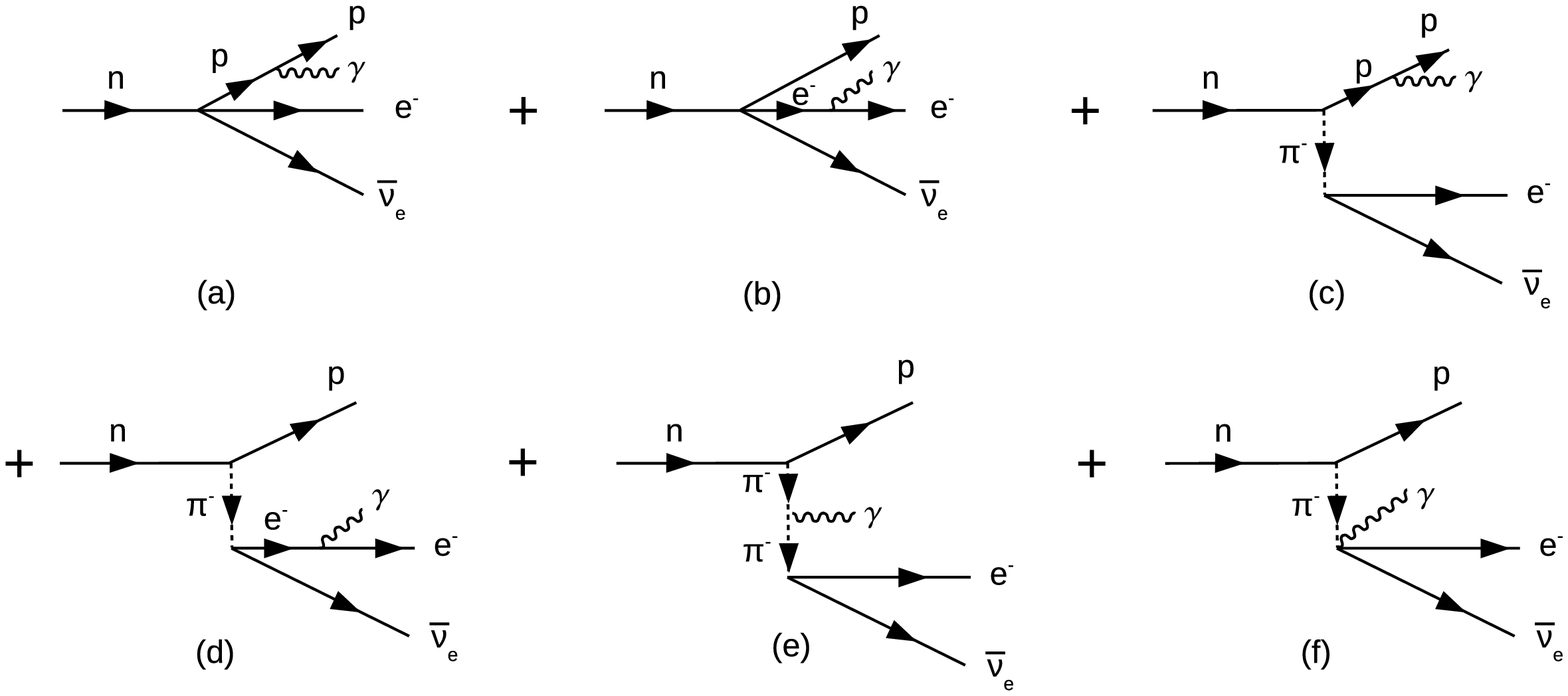}
  \caption{The Feynman diagrams the amplitude of the neutron radiative
    $\beta-$--decay within the combined quantum field theory including
    the standard $V - A$ effective theory of weak interactions, QED
    and the L$\sigma$M of strong low--energy interactions in the
    limit $m_{\sigma} \to \infty$, to leading order in the nucleon
    mass expansion and after renormalization.}
\label{fig:fig2}
\end{figure}

The Feynman diagrams of the amplitude of the neutron radiative
$\beta^-$--decay are shown in Fig.\,\ref{fig:fig2}.
The analytical expression of the amplitude of the neutron radiative
$\beta^-$--decay, defined by the Feynman diagrams in
Fig.\,\ref{fig:fig2} is given by \cite{Ivanov2018b}
\begin{eqnarray*}
\hspace{-0.3in}&& M(n \to p e^- \bar{\nu}_e \gamma)_{\lambda} = e
G_V\nonumber\\
\hspace{-0.3in}&&\times \Big\{\Big[\bar{u}_p(\vec{k}_p, \sigma_p)
  \gamma^{\mu}(1 - g_A\gamma^5) u_n(\vec{k}_n, \sigma_n)\Big]
\Big[\bar{u}_e(\vec{k}_e,\sigma_e)\,\frac{1}{2k_e\cdot k}\,Q_{e,
    \lambda}\,\gamma_{\mu} (1 - \gamma^5) v_{\nu}(\vec{k}_{\nu}, +
  \frac{1}{2})\Big]\nonumber\\
\hspace{-0.3in}&& - \Big[\bar{u}_p(\vec{k}_p,
  \sigma_p)\,Q_{p, \lambda} \,\frac{1}{2k_p \cdot k}\,\gamma^{\mu}(1 -
  g_A \gamma^5) u_n(\vec{k}_n,
  \sigma_n)\Big]\Big[\bar{u}_e(\vec{k}_e,\sigma_e) \gamma^{\mu} (1 -
  \gamma^5) v_{\nu}(\vec{k}_{\nu}, + \frac{1}{2})\Big]\nonumber\\
\hspace{-0.3in}&& + \frac{2 g_A m_N (q - k)_{\mu}}{m^2_{\pi} -
  (q - k)^2 - i 0}\,\Big[\bar{u}_p(\vec{k}_p, \sigma_p)\gamma^5
  u_n(\vec{k}_n, \sigma_n)\Big] \,\Big[\bar{u}_e(\vec{k}_e,\sigma_e)Q_{e,
    \lambda}\frac{1}{2 k_e\cdot k} \gamma^{\mu} (1 - \gamma^5)
  v_{\nu}(\vec{k}_{\nu}, +
  \frac{1}{2})\Big]\nonumber\\ 
\end{eqnarray*}
\begin{eqnarray}\label{eq:4}
\hspace{-0.3in}&& - \frac{2 g_A m_N
  q_{\mu}}{m^2_{\pi} - q^2 - i 0}\,\Big[\bar{u}_p(\vec{k}_p,
  \sigma_p)\,Q_{p, \lambda} \,\frac{1}{2k_p \cdot k}\,\gamma^5
  u_n(\vec{k}_n, \sigma_n)\Big]\Big[\bar{u}_e(\vec{k}_e,\sigma_e)
  \gamma^{\mu} (1 - \gamma^5) v_{\nu}(\vec{k}_{\nu}, +
  \frac{1}{2})\Big]\nonumber\\
\hspace{-0.3in}&& + \frac{2 g_A m_N q_{\mu}}{m^2_{\pi} - q^2 -
  i 0}\,\frac{(2 q - k)\cdot \varepsilon^*_{\lambda}(k)}{m^2_{\pi} -
  (q - k)^2 - i0}\,\Big[\bar{u}_p(\vec{k}_p, \sigma_p)\gamma^5
  u_n(\vec{k}_n, \sigma_n)\Big]\Big[\bar{u}_e(\vec{k}_e,\sigma_e)
  \gamma^{\mu} (1 - \gamma^5) v_{\nu}(\vec{k}_{\nu}, +
  \frac{1}{2})\Big]\nonumber\\
\hspace{-0.3in}&& + \frac{2
 g_A  m_N}{m^2_{\pi} - (q - k)^2 - i0}\,\Big[\bar{u}_p(\vec{k}_p,
  \sigma_p)\gamma^5 u_n(\vec{k}_n, \sigma_n)\Big]\,
\Big[\bar{u}_e(\vec{k}_e,\sigma_e)\hat{\varepsilon}^*_{\lambda}(k) (1
  - \gamma^5) v_{\nu}(\vec{k}_{\nu}, + \frac{1}{2})\Big]\Big\},
\end{eqnarray}
where $Q_{e,\lambda}$ and $Q_{p, \lambda}$ are given by
\cite{Ivanov2013,Ivanov2017,Ivanov2017b}
\begin{eqnarray}\label{eq:5}
\hspace{-0.3in}Q_{e, \lambda} = 2 \varepsilon^{*}_{\lambda}(k)\cdot k_e +
\hat{\varepsilon}^*_{\lambda}(k)\hat{k}\;,\; Q_{p,\lambda} = 2
\varepsilon^{*}_{\lambda}(k)\cdot k_p +
\hat{\varepsilon}^*_{\lambda}(k)\hat{k}.
\end{eqnarray}
Here $\varepsilon^*_{\lambda}(k)$ is the polarization vector of the
photon with the 4--momentum $k$ and in two polarization states
$\lambda = 1,2$, obeying the constraint $k\cdot
\varepsilon^*_{\lambda}(k) = 0$. For the derivation of Eq.(\ref{eq:1})
we have used the Dirac equations for the free proton and electron. The
first two terms in Eq.(\ref{eq:2}) are given by the Feynman diagrams
in Fig.\,\ref{fig:fig1}a and Fig.\,\ref{fig:fig1}b, whereas the last
four terms correspond to the contributions of the Feynman diagrams in
Fig.\,\ref{fig:fig2}c - Fig.\,\ref{fig:fig2}f, respectively. As has
been shown in \cite{Ivanov2018b} the contributions of the sum of the
Feynman diagrams in Fig.\,\ref{fig:fig1}a and Fig.\,\ref{fig:fig2}b
and the sum of Fig.\,\ref{fig:fig2}c - Fig.\,\ref{fig:fig2}f are
invariant independently under a gauge transformation of the photon
wave function $\varepsilon^*_{\lambda}(k) \to
\varepsilon^*_{\lambda}(k) + c\,k$, where $c$ is an arbitrary
constant. The contributions of strong low--energy interactions are
presented by the axial coupling constant $g_A$ and the one--pion-- and
two--pion--pole exchanges in the Feynman diagrams
Fig.\,\ref{fig:fig2}c - Fig.\,\ref{fig:fig2}f. In the amplitude of the
neutron radiative $\beta^-$--decay Eq.(\ref{eq:4}) we have omitted the
contribution of the weak magnetism. The contribution of this term
together with the proton recoil has been consistently taken into
account in the rate of the neutron radiative $\beta^-$-decay in
\cite{Ivanov2017}.

The contribution of the first two terms in Eq.(\ref{eq:4}) to the rate
of the neutron radiative $\beta^-$--decay, taken to leading order in
the large nucleon mass expansion with photon from the energy region
$\omega_{\rm min} \le \omega \le \omega_{\rm max}$, has been
calculated in \cite{Gaponov1996,Bernard2004,Ivanov2013,Ivanov2017}. It
takes the form \cite{Ivanov2013,Ivanov2017}
\begin{eqnarray}\label{eq:6}
\hspace{-0.3in}\lambda_{\beta\gamma}(\omega_{\rm max},\omega_{\rm
  min}) &=& \frac{\alpha}{\pi}\,(1 + 3
g^2_A)\,\frac{|G_V|^2}{\pi^3}\int^{\omega_{\rm
    max}}_{\omega_{\rm min}}\frac{d\omega}{\omega}\int^{E_0 -
  \omega}_{m_e}dE_e \,\sqrt{E^2_e - m^2_e}\,E_e\,F(E_e, Z = 1)\,(E_0 -
E_e - \omega)^2\nonumber\\
\hspace{-0.3in}&&\times\Big\{\Big(1 + \frac{\omega}{E_e} +
\frac{1}{2}\frac{\omega^2}{E^2_e}\Big)\,\Big[\frac{1}{\beta}\,{\ell
    n}\Big(\frac{1 + \beta}{1 - \beta}\Big) - 2\Big] +
\frac{\omega^2}{E^2_e}\Big\},
\end{eqnarray}
where $F(E_e, Z = 1)$ is the well--known relativistic Fermi function
of the Coulomb proton--electron final--state interaction, $E_e$ and
$\beta = \sqrt{E^2_e - m^2_e}/E_e$ are the energy and velocity of the
decay electron, $E_0 = (m^2_n - m^2_p + m^2_e)/2m_n = 1.2927\,{\rm
  MeV}$ is the end--point energy of the electron--energy spectrum
\cite{Ivanov2013}. In the experimental energy region $14\,{\rm keV}
\le \omega \le 782\,{\rm keV}$ \cite{Bales2016} the rate
Eq.(\ref{eq:6}) defines the branching ratio ${\rm Br}_{\beta\gamma} =
3.04\times 10^{-3}$, calculated for the neutron lifetime $\tau_n =
879.6(1.1)\,{\rm s}$ calculated in \cite{Ivanov2013} at $g_A =
1.2750(9)$ \cite{Abele2008} (see also \cite{Mund2013}). Such a
branching ratio does not contradict the experimental value ${\rm
  Br}_{\beta\gamma} = 3.35(16)\times 10^{-3}$ \cite{Bales2016} within
two standard deviations. The values of the neutron lifetime $\tau_n =
879.6(1.1)\,{\rm s}$ and axial coupling constant $g_A = 1.2750(9)$
agree well with recent values of the neutron lifetime $\tau^{(\rm
  favoured)}_n = 879.4(6)\,{\rm s}$ and axial coupling constant
$g^{(\rm favoured)}_A = 1.2755(11)$, which were recommended by
Czarnecki {\it et al.}  \cite{Sirlin2018} as {\it favoured}.

The contribution of these two terms, taken to leading order in the
large nucleon mass expansion, to the radiative corrections to the rate
of the neutron $\beta^-$--decay has been calculated in \cite{
  Berman1958}--\cite{Abers1968,Gudkov2006} (see also and
\cite{Ivanov2013}) within finite--photon mass regularization and takes
the form \cite{Ivanov2013}
\begin{eqnarray}\label{eq:7}
\hspace{-0.3in}\lambda_{\beta\gamma}(E_0,\mu) &=&
\frac{\alpha}{\pi}\,(1 + 3
g^2_A)\,\frac{|G_V|^2}{\pi^3}\int^{E_0}_{m_e}dE_e \,\sqrt{E^2_e -
  m^2_e}\,E_e\,F(E_e, Z = 1)\,(E_0 - E_e )^2\nonumber\\
\hspace{-0.3in}&&\times\Big\{\Big[2{\ell
n}\Big(\frac{2(E_0 - E_e)}{\mu}\Big) - 3 + \frac{2}{3}\,\frac{E_0 -
E_e}{E_e}\, \Big(1 + \frac{1}{8} \frac{E_0 - E_e}{E_e}
\Big)\Big]\Big[\frac{1}{2\beta}\,{\ell n}\Big(\frac{1 + \beta}{1 -
    \beta}\Big) - 1\Big] + 1 \nonumber\\
\hspace{-0.3in}&& + \frac{1}{12} \frac{(E_0 - E_e)^2}{E^2_e}+
\frac{1}{2\beta}\,{\ell n}\Big(\frac{1 + \beta}{1 - \beta}\Big) -
\frac{1}{4\beta}\,{\ell n}^2\Big(\frac{1 + \beta}{1 - \beta}\Big) -
\frac{1}{\beta}\,{\rm Li}_2\Big(\frac{2 \beta}{1 + \beta} \Big)\Big\},
\end{eqnarray}
where ${\rm Li}_2(z)$ is the Polylogarithmic function and $\mu$ is an
infinitesimal photon mass, which should be taken in the limit $\mu \to
0$ \cite{Berman1958}--\cite{Abers1968,Gudkov2006} (see also
\cite{Ivanov2013}). The rate of the neutron $\beta^-$--decay $n \to p
+ e^- + \bar{\nu}_e$, taking into account radiative corrections,
caused by virtual photon exchanges, calculated within the
finite--photon mass $\mu$ regularization, is given by (see Appendix D
of \cite{Ivanov2013})
\begin{eqnarray*}
\hspace{-0.3in}\lambda_{\beta}(E_0,\mu) &=& (1 + 3
g^2_A)\,\frac{|G_V|^2}{\pi^3}\int^{E_0}_{m_e}dE_e \,\sqrt{E^2_e -
  m^2_e}\,E_e\,F(E_e, Z = 1)\,(E_0 - E_e )^2\Big\{ 1+
\frac{\alpha}{\pi}\Big\{2 {\ell n}\Big(\frac{\mu}{
  m_e}\Big)\nonumber\\
\end{eqnarray*}
\begin{eqnarray}\label{eq:8}
\hspace{-0.3in}&&\times \Big[\frac{1}{2\beta}\,{\ell n}\Big(\frac{1 +
    \beta}{1 - \beta}\Big) - 1 \Big] + \frac{3}{2}{\ell
  n}\Big(\frac{m_p}{m_e}\Big) - \frac{11}{8} - \frac{1}{\beta}{\rm
  Li}_2\Big(\frac{2\beta}{1 + \beta}\Big) - \frac{1}{4\beta}\,{\ell
  n}^2\Big(\frac{1 + \beta}{1 - \beta}\Big) + \frac{\beta}{2}\,{\ell
  n}\Big(\frac{1 + \beta}{1 - \beta}\Big)\Big\}\Big\}.
\end{eqnarray}
Summing up the contributions of the rates $\lambda_{\beta}(E_0,\mu)$
and $\lambda_{\beta\gamma}(E_0,\mu)$ and taking the limit $\mu \to 0$
we arrive at the rate of the neutron decay
\begin{eqnarray}\label{eq:9}
\hspace{-0.3in}\lambda_n(E_0) =
(1 + 3
g^2_A)\,\frac{|G_V|^2}{\pi^3}\int^{E_0}_{m_e}dE_e \,\sqrt{E^2_e -
  m^2_e}\,E_e\,F(E_e, Z = 1)\,(E_0 - E_e )^2 \Big(1+
\frac{\alpha}{\pi}\,\bar{g}_n(E_e)\Big),
\end{eqnarray}
where the function $\bar{g}_n(E_e)$, defining the radiative corrections
to the neutron lifetime, is equal to \cite{Sirlin1967} (see also
Appendix D of \cite{Ivanov2013})
\begin{eqnarray}\label{eq:10}
\hspace{-0.3in}\bar{g}_n(E_e) &=& \frac{3}{2}\,{\ell
  n}\Big(\frac{m_p}{m_e}\Big) - \frac{3}{8} + 2\,\Big[\frac{1}{2\beta}
  \,{\ell n}\Big(\frac{1 + \beta}{1 - \beta}\Big) - 1\Big]\Big[{\ell
    n}\Big(\frac{2(E_0 - E_e)}{m_e}\Big) - \frac{3} {2} +
  \frac{1}{3}\,\frac{E_0 - E_e}{E_e}\Big]\nonumber\\
\hspace{-0.3in}&-& \frac{2}{\beta}{\rm Li}_2\Big(\frac{2\beta}{1 +
  \beta}\Big) + \frac{1}{2\beta}{\ell n}\Big(\frac{1 + \beta}{1 -
  \beta}\Big)\,\Big[(1 + \beta^2) + \frac{1}{12} \frac{(E_0 -
    E_e)^2}{E^2_e} - {\ell n}\Big(\frac{1 + \beta}{1 -
    \beta}\Big)\Big].
\end{eqnarray}
The contribution of the electroweak--boson exchanges together with QCD
corrections has been calculated by Czarnecki {\it et al.}
\cite{Sirlin2004} (see also \cite{Sirlin1986}). This defines the the
radiative corrections to the neutron lifetime, which are described by
the function $g_n(E_e) = \bar{g}_n(E_e) + C_{WZ}$, where $C_{WZ} =
10.249$ (see Appendix D of \cite{Ivanov2013}) is caused by
electroweak--boson exchanges and QCD corrections \cite{Sirlin2004}.

The infrared divergent contribution of hadronic structure of the
nucleon to the rate of the neutron radiative $\beta^-$--decay, induced
by the term $\sum_{\lambda =1,2}(\varepsilon^*_{\lambda}(k)\cdot
k_e)(\varepsilon_{\lambda}(k)\cdot k_e)/(k_e\cdot k)^2$ in the rate of
the neutron radiative $\beta^-$--decay
\cite{Ivanov2013,Ivanov2017,Ivanov2017b}, is equal to
\begin{eqnarray}\label{eq:11}
\hspace{-0.3in}&&\lambda^{(\rm h.s.)}_{\beta\gamma}(E_0,\mu) =
\frac{\alpha}{\pi}\,(1 + 3
g^2_A)\,\frac{|G_V|^2}{\pi^3}\int^{E_0}_{m_e}dE_e \,\sqrt{E^2_e -
  m^2_e}\,E_e\,F(E_e, Z = 1)\,(E_0 - E_e )^2 \,\Delta
\bar{g}^{(\rm h.s.)}_n(E_e,\mu).
\end{eqnarray}
Thus, the contribution of hadronic structure of the nucleon to the
radiative corrections of order $O(\alpha/\pi)$ to the neutron lifetime
is equal to
\begin{eqnarray}\label{eq:12}
  \Delta \bar{g}^{(\rm h.s.)}_n(E_e) &=& - \frac{2 g^2_A}{1 + 3
  g^2_A}\,\frac{m^2_e}{m^2_{\pi}}\Big\{2{\ell n}\Big(\frac{2 (E_0 -
  E_e)}{\mu}\Big)\Big[\frac{1}{2\beta}{\ell n}\Big(\frac{1 + \beta}{1
    - \beta}\Big) - 1\Big] + 1 + \frac{1}{2\beta}\,{\ell
  n}\Big(\frac{1 + \beta}{1 - \beta}\Big) - \frac{1}{4\beta}\,{\ell
  n}^2\Big(\frac{1 + \beta}{1 - \beta}\Big)\nonumber\\ &-&
\frac{1}{\beta}\,{\rm Li}_2\Big(\frac{2 \beta}{1 + \beta} \Big)\Big\}.
\end{eqnarray}
Relative to Sirlin's infrared divergent contribution Eq.(\ref{eq:7})
the order of this correction is of about $10^{-5}$. This confirms
validity and high confidence level of the contribution of hadronic
structure of the nucleon to the radiative corrections of the neutron
lifetime, calculated by Sirlin \cite{Sirlin1967} (see
Eq.(\ref{eq:10})), who proved a factorization of strong low--energy
and electromagnetic interactions to leading order in the large nucleon
mass expansion and, practically, dealt with structureless neutron and
proton.

The main problem of the infrared divergent correction Eq.(\ref{eq:12})
can be related only to a cancellation of such a divergence by the
contribution of virtual photon exchanges, that should be similar to
cancellation of the infrared divergence in Eq.(\ref{eq:7}) by the
infrared divergence in Eq.(\ref{eq:8}). An impossibility to cancel
such an infrared divergent correction Eq.(\ref{eq:12}), calculated to
leading order in the large nucleon mass expansion, by virtual photon
exchanges in the neutron $\beta^-$--decay should testify a certain
inconsistency of the approach.
Following \cite{Ivanov2018b} we may argue that a problem of possible
non--cancellation of the infrared divergence Eq.(\ref{eq:12}) by
virtual photon exchanges in the neutron $\beta^-$--decay can be related
to the use of the standard $V - A$ effective theory of weak
interactions. The point is that the effective $V - A$ vertex of
nucleon--lepton current--current interaction, accounting for both
baryonic and mesonic currents coupled to leptonic current, is not the
vertex of the combined quantum field theory including QED and
L$\sigma$M or any other theory of strong low--energy interactions. For
correct account for contributions of hadronic structure of the nucleon
to radiative corrections to the neutron $\beta^-$--decays one has to
use a combined quantum field theory including the Standard Electroweak
Model (SEM) and a theory of strong low--energy interactions (e.g. the
L$\sigma$M). However, first of all in our forthcoming publication we
are planning to perform an analysis of cancellation of the divergent
correction Eq.(\ref{eq:12}) by virtual photon exchanges in the neutron
$\beta^-$--decay within the combined quantum field theory described in
\cite{Ivanov2018b} and discussed in this paper.
\vspace{0.1in}

We thank Hartmut Abele for discussions stimulating the work under this
paper as a step towards the analysis of the SM corrections of order
$10^{-5}$ \cite{Ivanov2017a,Ivanov2017b,Ivanov2018a}. The work of
A. N. Ivanov was supported by the Austrian ``Fonds zur F\"orderung der
Wissenschaftlichen Forschung'' (FWF) under contracts P26781-N20 and
P26636-N20 and ``Deutsche F\"orderungsgemeinschaft'' (DFG) AB
128/5-2. The work of R. H\"ollwieser was supported by the Deutsche
Forschungsgemeinschaft in the SFB/TR 55. The work of M. Wellenzohn was
supported by the MA 23 (FH-Call 16) under the project ``Photonik -
Stiftungsprofessur f\"ur Lehre''.

\end{document}